\documentclass[12pt]{article}
\usepackage{graphicx}

\newcommand\pubnumber{SLAC--PUB--15224}
\newcommand\pubdate{August, 2012}

\def\SLAC{SLAC,
    Stanford University, Menlo Park, California 94025 USA}
\def\doeack{\footnote{Work supported by the US Department of Energy,
                     contract DE--AC02--76SF00515.}}

\def\Title#1{\begin{center} {\Large #1 } \end{center}}
\def\Author#1{\begin{center}{ \sc #1} \end{center}}
\def\Address#1{\begin{center}{ \it #1} \end{center}}

\newcommand\pubblock{\rightline{\begin{tabular}{l} \pubnumber\\
         \pubdate \end{tabular}}}
\newenvironment{Abstract}{\begin{quotation} \begin{center}
                       ABSTRACT
     \end{center}\bigskip  }{\end{quotation}}
\newenvironment{Presented}{\begin{quotation} \begin{center} 
             PRESENTED AT\end{center}\bigskip 
      \begin{center}\begin{large}}{\end{large}\end{center} \end{quotation}}

\def\Acknowledgements{\bigskip  \bigskip \begin{center} \begin{large}
             \bf ACKNOWLEDGEMENTS \end{large}\end{center}}

\def\beq{\begin{equation}}
\def\eeq#1{\label{#1}\end{equation}}
\def\eeqn{\end{equation}}

\newenvironment{Eqnarray}%
   {\arraycolsep 0.14em\begin{eqnarray}}{\end{eqnarray}}
\def\beqa{\begin{Eqnarray}}
\def\eeqa#1{\label{#1}\end{Eqnarray}}
\def\eeqan{\end{Eqnarray}}
\def\CR{\nonumber \\ }

\def\leqn#1{(\ref{#1})}

\let\bar=\overbar

\def\lsim{\mathrel{\raise.3ex\hbox{$<$\kern-.75em\lower1ex\hbox{$\sim$}}}}
\def\gsim{\mathrel{\raise.3ex\hbox{$>$\kern-.75em\lower1ex\hbox{$\sim$}}}}

\def\L{{\cal L}}

\def\O{{\cal O}}

\def\del{\partial}
\def\Dslash{\not{\hbox{\kern-4pt $D$}}}
\def\dslash{\not{\hbox{\kern-2pt $\del$}}}

\def\ee{e^+e^-}

\def\mw{m_W}

\def\msb{{\bar{\scriptsize M \kern -1pt S}}}

\def\drb{{\bar{\scriptsize D \kern -1pt R}}}

\def\eps{\epsilon}

\makeatletter
\def\section{\@startsection{section}{0}{\z@}{5.5ex plus .5ex minus
 1.5ex}{2.3ex plus .2ex}{\large\bf}}
\def\subsection{\@startsection{subsection}{1}{\z@}{3.5ex plus .5ex minus
 1.5ex}{1.3ex plus .2ex}{\normalsize\bf}}
\def\subsubsection{\@startsection{subsubsection}{2}{\z@}{-3.5ex plus
-1ex minus  -.2ex}{2.3ex plus .2ex}{\normalsize\sl}}

\renewcommand{\@makecaption}[2]{%
   \vskip 10pt
   \setbox\@tempboxa\hbox{\small #1: #2}
   \ifdim \wd\@tempboxa >\hsize     
       \small #1: #2\par          
     \else                        
       \hbox to\hsize{\hfil\box\@tempboxa\hfil}
   \fi}

 \def\citenum#1{{\def\@cite##1##2{##1}\cite{#1}}}
 
\newcount\@tempcntc
\def\@citex[#1]#2{\if@filesw\immediate\write\@auxout{\string\citation{#2}}\fi
  \@tempcnta\z@\@tempcntb\m@ne\def\@citea{}\@cite{\@for\@citeb:=#2\do
    {\@ifundefined
       {b@\@citeb}{\@citeo\@tempcntb\m@ne\@citea\def\@citea{,}{\bf ?}\@warning
       {Citation `\@citeb' on page \thepage \space undefined}}%
    {\setbox\z@\hbox{\global\@tempcntc0\csname b@\@citeb\endcsname\relax}%
     \ifnum\@tempcntc=\z@ \@citeo\@tempcntb\m@ne
       \@citea\def\@citea{,}\hbox{\csname b@\@citeb\endcsname}%
     \else
      \advance\@tempcntb\@ne
      \ifnum\@tempcntb=\@tempcntc
      \else\advance\@tempcntb\m@ne\@citeo
      \@tempcnta\@tempcntc\@tempcntb\@tempcntc\fi\fi}}\@citeo}{#1}}
\def\@citeo{\ifnum\@tempcnta>\@tempcntb\else\@citea\def\@citea{,}%
  \ifnum\@tempcnta=\@tempcntb\the\@tempcnta\else
  {\advance\@tempcnta\@ne\ifnum\@tempcnta=\@tempcntb \else\def\@citea{--}\fi
    \advance\@tempcnta\m@ne\the\@tempcnta\@citea\the\@tempcntb}\fi\fi}
\makeatother

\begin{document}
\begin{titlepage}
\pubblock

\vfill
\Title{Theoretical Summary Lecture for Higgs Hunting 2012}
\vfill
\Author{Michael E. Peskin\doeack}
\Address{\SLAC}
\vfill
\begin{Abstract}
In this lecture, I review some of the perspectives on the 
Higgs boson discussed at the Higgs Hunting 2012 Worshop and discuss the short-
and long-term aspects of Higgs physics.
\end{Abstract}
\vfill
\begin{Presented}
Higgs Hunting 2012 \\ LAL, Universite de Paris Sud, Orsay FRANCE \\
  July 18--20, 2012
\end{Presented}
\vfill
\newpage
\tableofcontents
\end{titlepage}

\def\thefootnote{\fnsymbol{footnote}}
\setcounter{footnote}{0}

\section{Introduction}

It is a rare pleasure to have attended the 2012 meeting of the Higgs
Hunters at LAL Orsay~\cite{HiggsHunting}.  The meeting was held two weeks after the remarkable
July 4 seminar at CERN that announced the discovery of a new boson 
of mass roughly 125 GeV, decaying to $\gamma\gamma$ and $ZZ^*$
~\cite{Higgsseminar}.  The euphoria
in the High Energy Physics community was still evident, and, I think it
will continue for some time.  It has been
a long time since I have had the pleasure of lecturing to an auditorium 
full of so many so happy people.

I myself am a bundle of emotions.  I am, all at the same time,
\begin{itemize}
\item   Awestruck
\item   Impatient
\item   Poised for the future
\end{itemize}
A discussion of the thoughts pulling me in these three directions gives as 
good a framework for discussing the current state of Higgs physics as any 
other.

\section{Awestruck}

I am awestruck by this discovery.

The discovery is not unexpected. The situation is quite the reverse:  
We have been waiting a long time 
for it.  I do not have space for a complete history, but here are 
crucial elements of the Higgs timeline:
\begin{itemize}
\item  {\bf 1967}:  Weinberg and Salam create their weak interaction
theory that requires the Higgs boson~\cite{Weinberg,Salam}.
\item {\bf 1975}:  Ellis, Gaillard, and Nanopoulos present the complete
 phenomenological profile of the Standard Model Higgs boson~\cite{EGN}.
\item {\bf 1976}:  Ioffe and Khoze and Bjorken discuss the $hZZ$ coupling as a means for 
discovering the Higgs boson~\cite{Ioffe,Bj}.
\item {\bf 1981}:   Okun declares the discovery of the Higgs to be ``Problem 
 Number 1'' in high-energy physics~\cite{Okun}.
\item{\bf 1982}:   The Snowmass 1982 workshop focuses high-energy physicists
 on the problem of electroweak symmetry breaking and the TeV scale~\cite{Snow}.
\item{\bf 1984}:   The ECFA-CERN Workshop on a Large Hadron Collider  initiates
the LHC~\cite{ECFACERN}.
\item{\bf 1987}:   Gunion, Kane, and Wudka discuss $h\to \gamma\gamma$
  and $h\to ZZ^*$
   as means for discovering the Higgs boson~\cite{GKW}.
\item{\bf 1993}:  The Superconducting SuperCollider is cancelled, dealing
  a setback to Higgs hunters.
 \item{\bf 1995}:  The discovery of the top quark sharpens the precision
   electroweak implications for the Higgs boson, predicting a low
   value for the Higgs mass~\cite{EWWG}
 \item{\bf 2000}:   LEP runs at 209~GeV in its last days, giving hope but not 
success in the search for the Higgs boson~\cite{LEP}
\end{itemize}
We can add July 4, 2012, to this history.

It is not only the time scale of the Higgs boson search that is impressive.
Those of us who scribble equations for a living are humbled by the enormous
effort it takes to find out whether those equations are relevant to the 
real world.  I felt this already in the days when physicists explored 
energies of just a few GeV in teams of thirty to fifty.  It is even more 
awe-inspiring to watch the ATLAS and CMS experiments pursue their 
analyses.   The enormous scale of the endeavor is measured in some 
obvious ways---the 27~km scale of the accelerator, the highest energy 
particles accelerated by man, the world's largest cryogenic system, the
5-story-high particle detectors, the 3000 physicists contributing to each
publication.  But there is more.   I would like to highlight three more 
items.

First, the discovery of the Higgs boson is the world's hardest data problem.
Many scientists and engineers today tout their analyses of Big Data.  But 
nothing is bigger than this.  The Higgs boson appears, in the decay modes
used for the discovery, in fewer than 1 in $10^{12}$ proton-proton collisions.
To search for the Higgs, ATLAS and CMS push out an enormous stream of 
raw data,  100 Tb/sec.  The permanent databases of these experiments are 
tens of Pb.  It is dangerously close to  true that there are not enough 
computers or 
human brains in the 
world physics community to understand this data, so a crucial part of the 
logistics of the experiments
 is the  global sharing and analysis of these huge databases.  The day
before the workshop, the Herald Tribune reminded us that, in July 1962, the
Telstar satellite began the global information revolution with the
first 
television pictures broadcast live
across the Atlantic~\cite{Telstar}.  Fifty years later, it is 
our community that is at the cutting edge.

Second, the LHC and the ATLAS and CMS experiments have relied on the intense 
commitment of scientists and laboratories over the past 25 years.  
At the workshop, Daniel Denegri told a part of this story~\cite{Denegri}.
Most moving to me are the stories of the 
people who began with ATLAS and CMS in the 
mid-1980's as young postdocs and have devoted their whole careers to 
preparing the infrastructure for this discovery.   The list of these 
people includes recent spokesmen of the collaborations---Jim Virdee and 
Fabiola Gianotti---but there are many others to thank.  These include our 
LAL hosts Louis Fayard and Daniel Fournier, who played a key role in enabling
the ATLAS electromagnetic calorimetry to see the Higgs decay to 
$\gamma\gamma$.
I must also point to the amazing institutional continuity and persistence of 
CERN---across 6 DG terms---to realize the LHC project, and the continued 
support of the taxpayers of Europe.  I know how difficult this is; we 
tried a similar effort
in the United States, but we could not make it work.

Finally, I am impressed by the enormous effort in QCD calculation that
has been carried out over decades to produce reliable theoretical predictions
for signal and background in Higgs boson searches.  These were reviewed at the
workshop by Robert Harlander, who is one of the important contributors
to this effort~\cite{Harlander,GrazzReview}.   These calculations are among
the most difficult that have ever been done in physics.  They require not 
only persistence but also great creativity.  

These amazing achievements brought us to the July 4 discovery.   After 
July 4, we find ourselves in a new era of particle physics.  Many 
questions that we had before have become irrelevant.  Other questions 
need to be posed anew.  And, some questions whose importance we could not
see before the discovery have now become central. 

\section{Impatient}

  Let us, then, turn to
the discussion of what we know and what we need to know.  I am impatient
to know more about this particle.  I sketch below a framework for organizing
the questions.

\subsection{Is it the Higgs boson?}

The first question is: Do we actually have grounds to call the newly 
discovered particle the Higgs boson?  The issue is obviously not 
settled.  However, there is an argument that is surprisingly strong 
for the current early stage in the study of this particle.  

The fact that the particle decays to $\gamma\gamma$ implies tha the
particle must be a boson and, by the Landau-Yang 
theorem~\cite{Landau,Yang},
cannot be spin 1.  Then we already know that it is a new type
of elementary particle, one different from all other particles of 
the Standard Model.   It is very difficult to exclude spin 2 and 
higher, because these theories can mimic spin 0, but certainly spin 0
is the simplest choice.

Many types of spin 0 particles can couple to $\gamma\gamma$ and to $gg$
through loop diagrams.  However, couplings to $WW$ are more restricted.
The Standard Model Higgs boson has an order 1 coupling to $WW$ generated
from its gauge-invariant kinetic term.   Starting from
\beq
      \L =  | D_\mu \varphi(x) |^2 \ ,
\eeq{starthere}
we assume that the field $\varphi$ acquires a vacuum expectation value $v$.
Let $h(x)$ be the field that corresponds to a space-time variation of this
vacuum expectation value. 
Then \leqn{starthere} becomes
\beqa
     \L &=& {g^2\over 4} (v + h(x))^2 W^+_\mu W^{-\mu} \CR
        &=&  \mw^2  W^+_\mu W^{-\mu} + {2\mw^2\over v} h \  W^+_\mu W^{-\mu}
           + \cdots
\eeqa{findWW}
This argument generates a similar Higgs coupling to $ZZ$ with strength $2m_Z^2/v$.

A spin 0 field that does not have a vacuum expectation value can also 
couple to $WW$ and $ZZ$ in a manner symmetric under $SU(2)\times U(1)$
through dimension 5 operators involving the 
$W$ and $Z$ field strengths.  In a weak-coupling theory, these operators
are generated by loops and so are suppressed by a power of $\alpha$.
These terms have the form
\beq
\L = A {\alpha\over 4\pi}{1\over M} h \ F_{\mu\nu} F^{\mu\nu} + 
     B {\alpha\over 4\pi}{1\over M} h \ \eps_{\mu\nu\lambda\sigma}
F^{\mu\nu} F^{\lambda\sigma}
\eeq{ABterm}
We see the new particle 
coupling to $WW$ and $ZZ$ with a strength similar to that predicted in 
the Standard Model, rather than two orders of magnitude smaller.
From the choice of vertices above, this is 
{\it prima facie} evidence that the new particle is a CP even spin 0
field with a vacuum expectation value that breaks $SU(2)\times U(1)$.
This is exactly what we call a ``Higgs boson''.

This argument is hardly airtight.  Vertices of the type \leqn{ABterm} with 
order 1 coefficients can be generated in strong-coupling theories of 
TeV scale physics.  Spin 2 particles can have direct non-derivative 
couplings to $WW$ and $ZZ$.

\begin{figure}
\begin{center}
\includegraphics[width=2.2in]{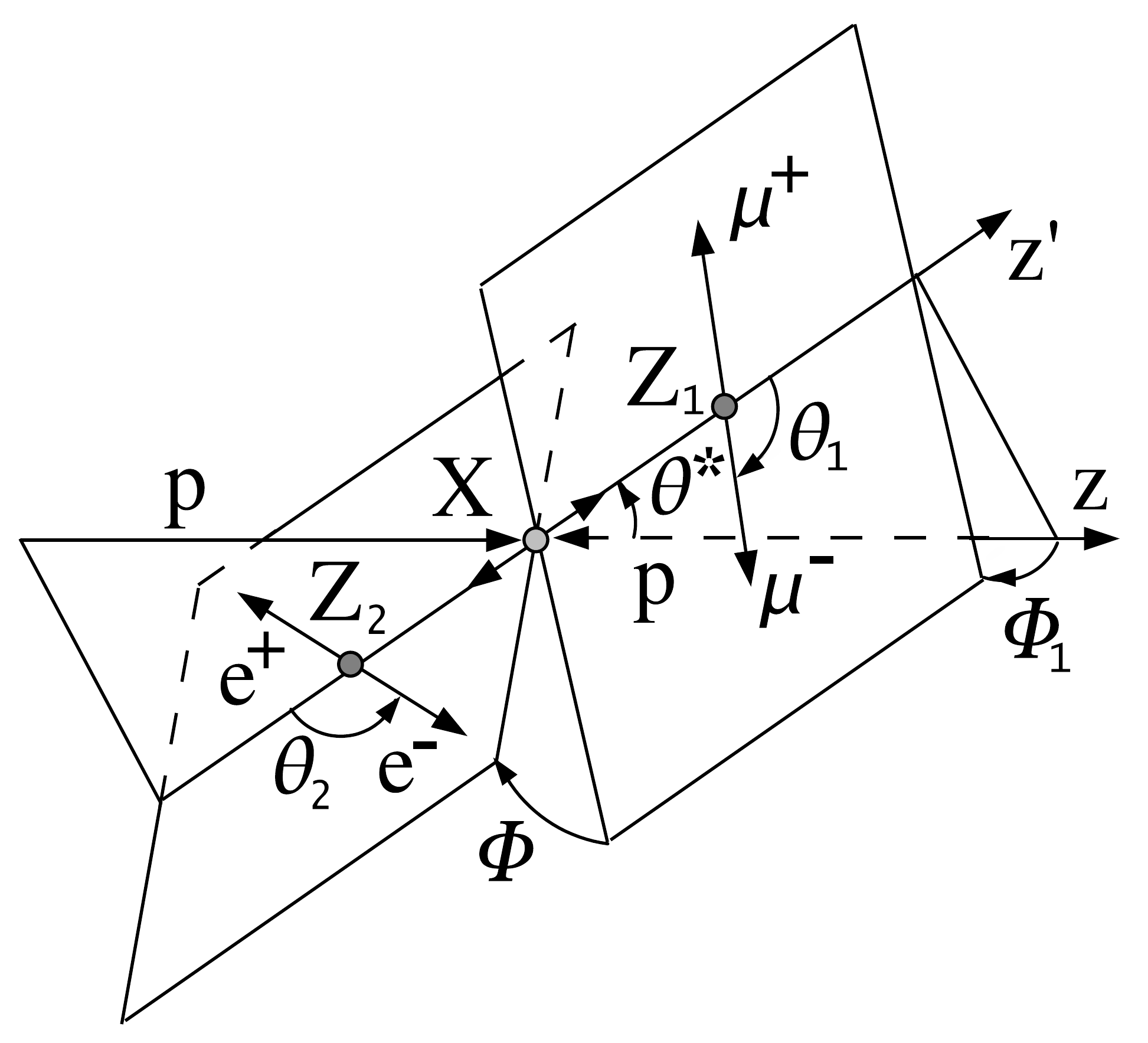}
\caption{Angles used in the spin analysis of the new particle in its
   4-lepton final state, from \cite{Gritsan}. }
\label{fig:ZZspin}
\end{center}
\end{figure}

However, we can find further support for the Higgs field interpretation
by studying the spin correlations in the process~\cite{Gritsan,Lykken}
\beq
        pp \to ZZ^* \to    \ell^+\ell^-\ell^+\ell^-
\eeq{pptoZZ}
The reconstruction of the new particle in the four lepton final state
allows us to measure the five angles shown in Fig.~\ref{fig:ZZspin}.  The
angle $\theta^*$ is sensitive to the production dynamics and discriminates
production of an $s$-channel resonance from the background process
$q\bar q \to ZZ$.   However, the angles $\theta_1$, $\theta_2$, and 
$\Phi_1-\Phi_2$ are sensitive to the decay dynamics.  In particular, they 
distinguish the vertex in \leqn{findWW}, in which the two $Z$s are 
dominantly longitudinally polarized, from \leqn{ABterm}, in which the 
two $Z$s are transversely polarized.  This angular analysis was described
at the workshop in the talk of Baffioni on the CMS observation of the 
new particle in $ZZ^*$~\cite{Baffioni}.   This angular
analysis already distinguishes the scalar and pseudoscalar cases at 
about 1 sigma.   Baffioni reported that 3 sigma separation is possible
with 30 fb$^{-1}$ at 8 TeV.

From here on, I will call the new particle ``the Higgs Boson'' without
further apology.

We must still find out whether this particle has the properties
predicted for the Higgs boson in the Standard Model.   The 
Standard Model insists that the Higgs boson is the unique source of
mass for all quarks, leptons, and gauge bosons. This  implies that the
couplings of the boson to all quarks, leptons, and gauge boson are
 precisely in the ratio of their masses, up to simple factors
 reflecting the particle spins. 
It is really so?

The mass of 125~GeV makes the Standard Model Higgs boson exceptionally
hard to find.  However, once we have found the particle, this special
mass
confers an advantage.   At this mass, the Standard Model Higgs boson
has a large number of decay channels with substantial branching
fractions available for study. 
 As Fabiola Gianotti put it in her July 4 lecture:  ``Thank
you, Nature."

Mele reviewed the phenomenology of the Standard Model Higgs boson at
the mass of 126.~GeV, referring to it properties as ``the new set of
Standard Model reference parameters''~\cite{Mele,Djouadi}.  The predicted width of the
boson is 4.2~MeV.   The major branching fractions are:
\begin{quote}
\begin{tabular}{crccrccr}
  $b\bar b$    &     56\% & \qquad \qquad   &  $\tau^+\tau^-$  &  6.2\%    
  &  \qquad \qquad & $\gamma\gamma$  &   0.23\% \\
$WW^*$       &     23\%   & \qquad \qquad &   $ZZ^*$   &   2.9\%  &   \qquad  \qquad& $\gamma Z$ &
0.16\% \\
$gg$    &    8.5\%    & \qquad  \qquad  &     $c\bar c$   &  2.8\%   & \qquad  \qquad  &  $\mu^+\mu^-$
&  0.02\% \\ 
\end{tabular}
 \end{quote}  
For all of these modes except $c\bar c$, there is a
strategy to observe the decay at the LHC.

Our understanding of the new boson will proceed in stages.  I foresee three stages:
\begin{itemize}
\item Are the major decay modes present?
\item Is the boson Standard Model-like, or not?
\item Are there small deviations from the Standard Model predictions?
\end{itemize}
Let's discuss these questions one by one.

\subsection{Are the major decay modes present?}

Already by the time of this meeting, many of the key qualitative
properties of a Standard Model Higgs boson are being
confirmed. Further information was provided after the conference
 in the papers submitted by
ATLAS~\cite{ATLASpaper}, CMS~\cite{CMSpaper}, and the Tevatron
experiments~\cite{TeVpaper}.
Here
is a list of the most important nine items, and the current status of
each:
\begin{enumerate}
\item $\gamma\gamma$ decay mode:   Observed (4.5  $\sigma$ in ATLAS,
4.1  $\sigma$ in CMS) .
\item  $ZZ^*$ decay mode:   Observed   (3.6  $\sigma$ in ATLAS,
3.2  $\sigma$ in CMS) .
\item  $WW^*$ decay mode:   Observed   (2.8  $\sigma$ in ATLAS,
1.6  $\sigma$ in CMS) .
\item $b\bar b$ decay mode.   So far, this is seen only by the
  Tevatron experiments, at 2.8 $\sigma$ in the CDF/D\O\  combination.
  CMS seems to be making good progress toward the observation at the
  LHC.~\cite{CMSbbbar}
\item $\tau\tau$ decay mode:  This is not yet observed; CMS reports a
  deficit with respect to the expectation.
\item  Spin-Parity:   As noted above, there is a preliminary
  indication from the CMS spin analysis of the $ZZ^*$  decay.
\item Gluon Fusion production mode:  This is the dominant production 
model for the observation of the boson in $\gamma\gamma$.
\item Vector Boson Fusion production mode:   ATLAS claims that the 
rate of Vector Boson Fusion production and $\gamma\gamma$ observation
is nonzero at 2.7 $\sigma$  significance.  CMS claims 3.5 $\sigma$
significance for $\gamma\gamma$ production with a ``VBF tag", a weaker 
statement.
\item Higgsstrahlung production mode: Seen at the Tevatron
  only, in the $b\bar b$ final state listed above.
\end{enumerate}
This is quite an impressive scorecard.  It is very likely that all of
the issues listed here will be settled, at the yes/no level, with the
full 2012 data set from the LHC.

\subsection{Is the Higgs Standard Model-like, or not?}

\begin{figure}
\begin{center}
\includegraphics[width=6.0in]{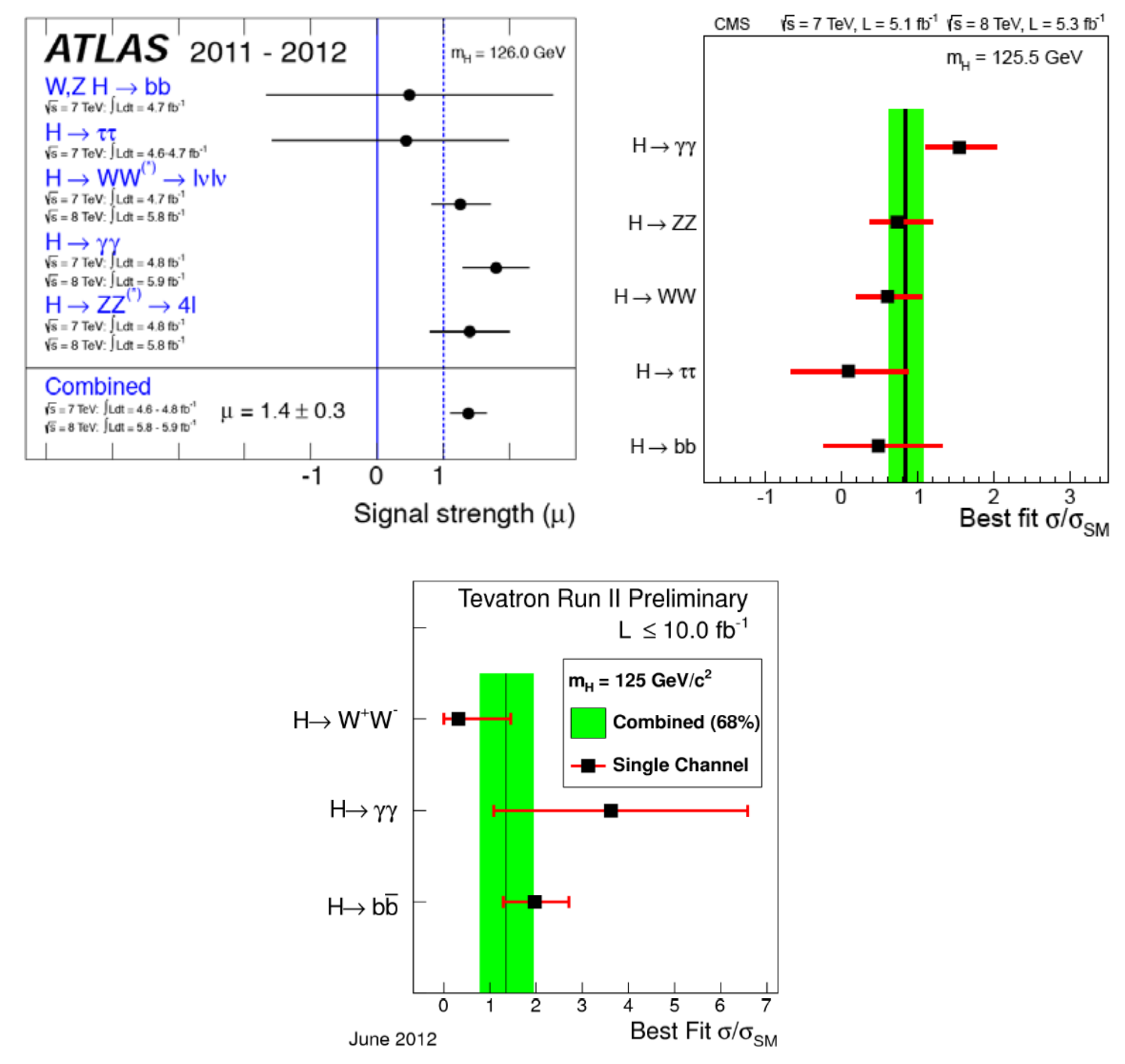}
\caption{Measured relative signal strength $\mu$ in many channels of
  the ATLAS, CMS, and Tevatron Higgs searches:  (a)  from ATLAS~\cite{ATLASpaper}, (b)
  from CMS~\cite{CMSpaper}, (c) from the CDF and D\O \ combination~\cite{Haley}.  }
\label{fig:mus}
\end{center}
\end{figure}
 
 There is much interest now in parsing the deviations from the Higgs
boson production and decay  rates predicted by the  Standard Model.
These rates are determined by a combination of Higgs properties, as I will discuss
in a moment.  A measurement of the rate for production of the
Higgs
boson at the LHC gives the
 {\it relative signal strength} $\mu$, defined by 
\beq
     \mu =    \sigma \cdot  BR /  (\sigma\cdot BR)|_{SM} \ ,
\eeq{mudef}
where $\sigma$ is the Higgs production cross section in the
measurement under consideration and $BR$ is the branching ratio of the
Higgs into the final state observed in the analysis. Here and below,
$SM$ denotes the Standard Model prediction.  The production
cross section will in general be a combination of the Gluon Fusion, 
Vector Boson Fusion, and other elementary cross sections, as defined
by the particular set of cuts used in the measurement.

The ATLAS and CMS experiments and the Tevatron experiments have
presented values of $\mu$ for a variety of final states and cross
section tags.  These are shown in Fig.~\ref{fig:mus}.    The fact that
the central value of $\mu$ is close to 2 in several channels, in
particular, in the LHC $\gamma\gamma$ signal and 
$\mu$  in several channels, in particular, for the LHC
$\gamma\gamma$ signals and the Tevatron $b\bar b$ signal, has excited
much interest.  However, we are still at an early stage in the study
of the Higgs, and these large signals are consistent with the expected
size of  fluctuations. 

The analysis of $\mu$ deviations is very much fun for theorists.
There are many interesting model-building solutions that give order 1
modifications of the Higgs boson signal strengths. 
 These typically involve new particles with masses of the order
of 200~GeV or below~\cite{lowLow}. A nontrivial part of the game is to suggest new
particles that are not excluded by the LHC experiments.  Possible new
particles influencing the Higgs rates include new bosons from an
extended Higgs sector~\cite{Uli,Gunion}, new color-singlet matter 
particles such as the
tau slepton~\cite{Marcella}, or  new colored particles such as light
top squarks that are stealthy at the LHC~\cite{Stop}.  Strong
interactions in the Higgs sector can also influence the Higgs signal
strengths;  a compositeness scale close to 1~TeV is 
required for a large effect~\cite{Grojean,Neubert}.  Carena gave examples of many of
these scenarios in her talk at the workshop~\cite{Carena}. 

There are many groups now that fit the measured signal strengths to
look for insight.  Some of these fits were reviewed at the workshop by 
Espinosa~\cite{Espinosa}.   At the moment, fits to the current
measurements tend to be 2-parameter fits under specific model
hypotheses.  They give insight if the particular scheme assumed for
modifying the Standard Model is correct.  

It is important to realize, though, that analyses of the Higgs
properties in terms of a small number of parameters bring in
assumptions that might well be incorrect.  It is easy to construct
models that tweak individual Higgs couplings away from their Standard
Model values without affecting other couplings.  Models with two Higgs 
doublets can modify the Higgs couplings to up-type quarks, down-type
quarks, or leptons without changing the couplings to other matter
particles.  Introduction of new particles that appear in loops can
modify the Higgs couplings to $\gamma\gamma$ and $gg$ while having
small effects on the other couplings.  

This means that  a deviation of a signal strength
$\mu$ from 1 gives ambiguous information, pointing in several different
directions.   A given $\mu $ parameter refers to a production channel
$A\bar A \to h$ (where $A\bar A$ might, for example, be $gg$ or 
$WW$) and a decay channel $h \to B\bar B$.  Since $\mu$
contains the branching ratio, the total width of the Higgs also
enters.  In all
\beq
           \mu(A\bar A \to h \to B\bar B) =  {\Gamma(A)\Gamma(B)\over
             \Gamma_T} /  SM  \ ,
\eeq{mucompute}
where $\Gamma(A)$ is the partial width for Higgs decay to $A\bar A$,
$\Gamma_T$ is the total width of the Higgs, and $SM$ is the Standard
Model value of the numerator.   An excess in
the rate for Higgs production by Gluon Fusion and observation in
$\gamma\gamma$
might be due to: 
\begin{itemize}
\item an enhancement of $\Gamma(\gamma)$
\item an enhancement of $\Gamma(g)$
\item a suppression of $\Gamma(b)$, the dominant component of
  $\Gamma_T$.
\end{itemize}
or any combination of these effects.  A small enhancement could be due
to a suppression of $\Gamma(W)$. 

At our current state of experimental uncertainty, global fits to the Higgs
couplings that allow all of these deviations to fluctuate
independently
are unstable.  We
will need more data, and, probably, more data than the LHC will
provide in 2012, to resolve the ambiguities.

However, if this problem will not be solved this year, there are good
prospects for a qualitative understanding of the Higgs properties from
the LHC run at 14~TeV.   In principle, we would like to make global
fits to the rates of Higgs production and decay processes that include 
the  couplings to all of the Higgs decay modes listed at the end of
Section 3.1, plus the Higgs couplings to $t\bar t$ and to invisible
decay  product.
 This is an ambitious goal, but, with the help of a
weak theoretical assumption, it is within the reach of the LHC.

There are two problems on the path to getting the inputs required for
such global fits.  The first is that the dominant decay mode of a
Standard Model Higgs boson of mass 125~GeV, the decay $h\to b\bar b$,
is very difficult to observe at the LHC.  The problem is the enormous 
background from QCD production of $b\bar b$ pairs, at the  $\mu$b
level compared to the pb level for Higgs production.  To overcome this
problem, it is necessary to observe the Higgs in particular
characteristic reactions, especially, in associated production with
$W$, $Z$, or $t\bar t$.  This does not solve the problem, however.
A reaction that has a much higher cross section than $pp\to
Wh$ is $pp\to Wg$, with subsequent gluon splitting to $b\bar b$. 
Even after a cut on mass of the $b\bar b$ system around the known Higgs mass,
the Higgs signal is submerged in background.

Recently, a solution to this problem was proposed by Butterworth,
Davison, Rubin, and Salam~\cite{BDRS}.  These authors begin from the
idea that, if 
the Higgs is highly boosted, the $b$ and $\bar b$ jets are merged into
a single anti-$k_T$ jet.  They then note that the internal structure
of this jet is different from that of a gluon jet with splitting to
$b\bar b$.   The Higgs jet has fewer soft subjets, a consequence of
its color-singlet rather than color-octet origin, and its major two
component subjects share their energy more equally, a consequence of
its origin as a massive particle.  With these features in mind,
Butterworth {\it et al.} devised a ``jet grooming'' strategy that removes
the gluon background.   Plehn, Salam, and  Spannowsky proposed a
similar grooming strategy for the measurement of the cross section for
$pp\to t\bar t h$~\cite{PSS}.  The study of boosted objects and jet
grooming is  now of interest for many applicationsto LHC 
physics; the subject has
recently been reviewed in~\cite{Boost} and in Salam's presentation to
this workshop~\cite{GSalam}.

The second problem is to control the branching ratio of the Higgs to 
final states
that are not visible to the LHC experiments.  An example is the decay 
$h\to c\bar c$, for which, currently, there is no strategy for its
observation at the LHC.   This requires a theoretical argument that 
this branching ratio cannot be large.  Such an argument can be made by
using the idea that, if there are many fields with vacuum expectation
values that contribute to the $W$ and $Z$ masses, each makes a
positive contribution, and these sum to the observed vector boson
masses.  This idea implies the inequalities
\beq
    \Gamma(W) \leq \Gamma(W)|_{SM} \qquad 
 \Gamma(Z) \leq \Gamma(Z)|_{SM} 
\eeq{SMineq}
Gunion, Haber, and Wudka have shown that these inequalities are
generally true in models with no CP violation in the Higgs sector and
no doubly charged Higgs bosons~\cite{GHW}.  D\"uhrssen and
collaborators introduced the use of these inequalities in Higgs
parameter fitting in \cite{Duhrssen}.  Using this assumption, it
is possible to make a controlled fit to LHC data with the full set of free
parameters listed above. 

\begin{figure}
\begin{center}
\includegraphics[width=6.0in]{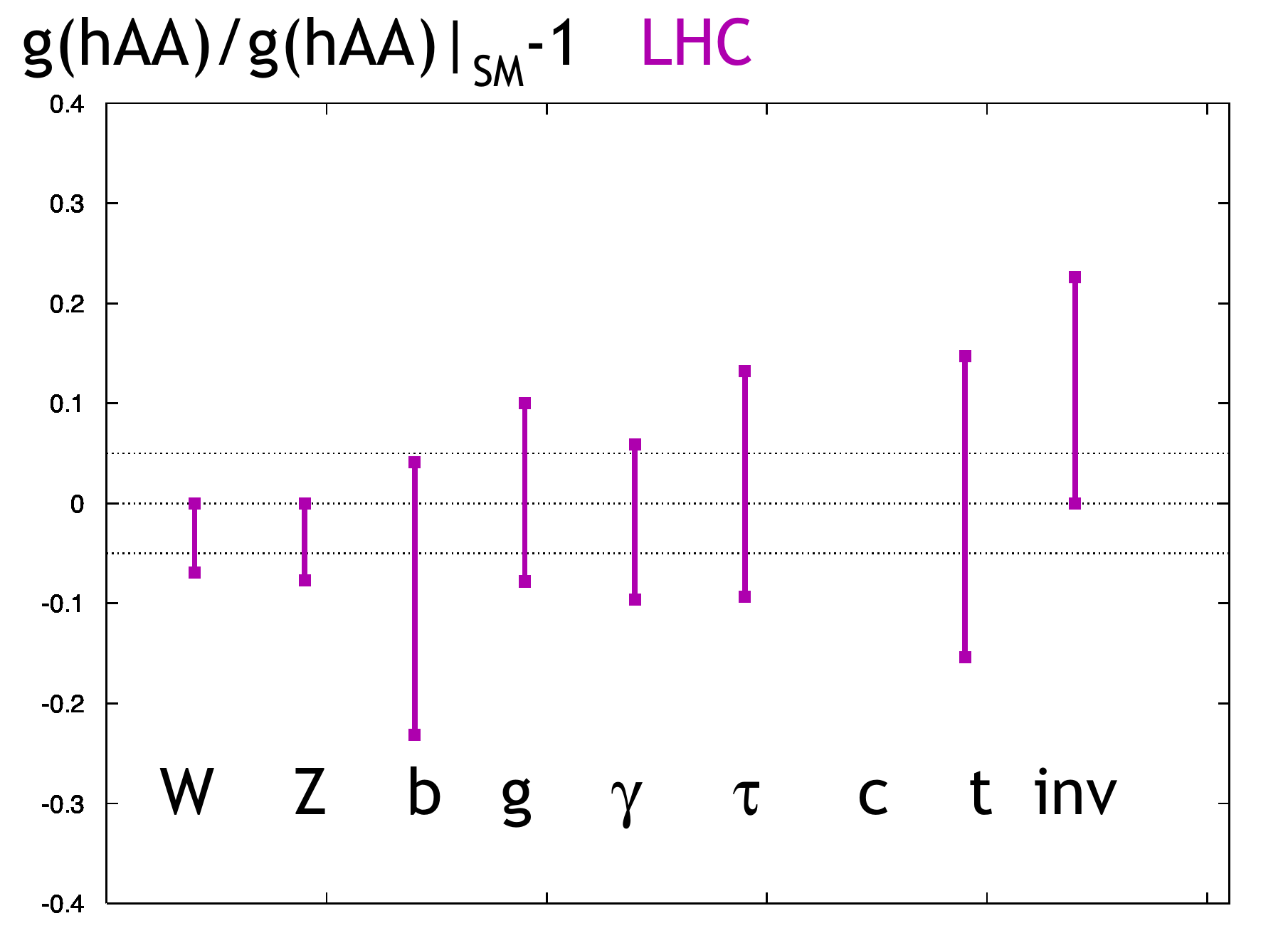}
\caption{Estimates of the accuracy that can be achieved in Higgs
  coupling measurements using a model-independent fit to LHC
  measurements with a 300~fb$^{-1}$ data set, from \cite{MyHiggs}.
  The estimates are given as a fraction of the predicted Standard
  Model value for the Higgs coupling constants.  The indicated
  horizontal lines
  represent 5\% deviations.  For the invisible
  Higgs decay,
 the quantity plotted is the square root of the branching fraction.}
\label{fig:HiggsLHC}
\end{center}
\end{figure}

Following this idea, I made an estimate of the accuracy to which the
LHC results available by the end of the decade would constrain the
Higgs couplings in such a model-independent fit~\cite{MyHiggs}. 
The analysis makes strong use of the work of D\"uhrssen and the 
Heidelberg group~\cite{Duhrssen,Duhrssenthesis}.  The analysis also 
takes account of new projections prepared by ATLAS and CMS for the 
European Strategy Study~\cite{ATLASESS,CMSESS}. The 
results are shown in Fig.~\ref{fig:HiggsLHC}.   Klute {\it et
  al.}~\cite{Klute} have done a similar study and have reached similar
conclusions.  My analysis is less sophisticated but, I hope, more
transparent in terms of the assumptions used.   Neither study makes
use of the improved knowledge of the ATLAS and CMS detector
capabilities that has been obtained through actual data-taking and
analysis.   I hope that the ATLAS and CMS collaborations will soon study
this question and report improved estimates of
 their prospects for Higgs boson measurements.

The results are quite striking.  The analysis sets a high standard by 
which to measure the LHC capabilities.  The conclusion is that the LHC
experiments are capable of being evaluated by this standard, and that 
these experiments---with large data sets of about 300~fb$^{-1}$---will give 
accurate assessments of the individual Higgs boson couplings, with errors
at the level 
of 10--20\%  within a 
model-independent analysis.  This will settle the question of whether or
not the newly discovered boson has properties close to those of the 
Standard Model Higgs boson.

And, yet, this level of accuracy is not good enough.

\subsection{Are there small deviations from the Standard Model?}

Must we care about Higgs boson coupling measurements below the 10\% level? 
In fact, measurements of even higher accuracy are likely to provide an 
essential part of the Higgs boson story.

There are two important points to be made here.

First, although the Higgs boson may turn out to look Standard Model-like
by the standards just described, and although it is possible that no new
particles will be discovered at the LHC in its first sample of a few
hundred fb$^{-1}$, we cannot give up on the idea that there is new physics
beyond the Standard Model at the TeV energy scale.  It may turn out that
the precision study of the Higgs boson is our best route to uncovering
evidence of that new physics.

Much has been said about the incompleteness of the Standard Model and its
inadequacy as a model of electroweak symmetry breaking.  I have little to 
add on this point except to put the issue bluntly.  In the Standard Model,
the {\it complete explanation} for the spontaneous breaking of the 
$SU(2)\times U(1)$ electroweak gauge symmetry is the following:  The theory
has a parameters $\mu^2$.  Electroweak symmetry is broken if 
\beq
       \mu^2(\mbox{TeV}) < 0 \ .
\eeq{musqless}
That's it.   Since $\mu^2$ is additively renormalized, there is no natural
distinction between positive and negative values of $\mu^2$.   As physicists,
we should be ashamed of ourselves if we are satisfied with this.

The second point is less widely recognized. Many classes of models of 
electroweak symmetry breaking contain a light Higgs boson similar to the
Higgs boson of the Standard Model.  Examples include supersymmetry, Little
Higgs models, and Randall-Sundrum extra-dimensional models.  After 
July 4, any model that does not predict a light Higgs that is the major 
source of electroweak symmetry breaking is at a severe disadvantage.  At
the moment, is it still true that certain models without a light Higgs 
are not excluded~\cite{technidil,dil}, but they will be 
in deep trouble if the measurements described in Section 3.2, and available
this year, meet the Standard Model expectations.

In Section 3.3, I made reference to many models that predicted order 1
deviations of the Higgs boson couplings from the Standard Model predictions.
Most of these models have a common feature of requiring new particles with
masses of the order of 200~GeV or below.  Those models that  modify the
Higgs boson couplings through strong interaction effects in the 
electroweak sector require large perturbations not only in the Higgs 
couplings but also in the top quark and $W$ boson couplings.  If these
particles or effects are not found, what then?

The more typical prediction of new physics models is that the new physics 
effects on Higgs boson couplings are quite small.  In the 1990's, Howard
Haber discussed this conclusion in very general terms in~\cite{Haber}.
Haber defined the ``Decoupling Limit'' of a new physics model in which the
Higgs boson is light  but other new particles are  heavy, at masses of 
1~TeV or above.  In this situation, the fields associated with
 the new particles can be integrated out of the effective Hamiltonian 
describing Higgs physics.  The effects of these particles is then present
only in higher-dimension operators whose coefficients are of the  order 
of 
\beq
         m_h^2/M^2 \quad \mbox{or} \quad m_t^2/M^2 \ , 
\eeq{decouple}
where $M$ is the mass of the new particles.

Here are some examples of corrections to the Higgs couplings in specific 
models of new physics.  More examples, and further discussion of the 
Decoupling Limit, can be found in the recent paper of Gupta, Rzehak, and
Wells~\cite{GRW}.

In supersymmetric models, it is necessary
that there are at least two Higgs 
doublet fields.  This gives rise to corrections to the Higgs couplings at
tree level.  The typical size of the corrections to the $h\tau\tau$ 
coupling is~\cite{taucoupling}
\beq
      g(\tau)/SM = 1 +  10\% \left({\mbox{400\ GeV}\over m_A}\right)^2 \ , 
\eeq{SUSYone}
where $m_A$ is the mass of the heavy $A^0$ Higgs boson.  In models with
large $\tan\beta$, the $hb\bar b$ coupling receives additional corrections
from loop diagrams, estimated as~\cite{bcoupling}
\beq
    g(b)/SM = g(\tau)/SM + (1 - 3)\% \ .
\eeq{SUSYtwo}

In Composite Higgs models, the Higgs bosons are effective Goldstone
boson fields. The Higgs couplings receive corrections sized by the 
scale of the Goldstone boson decay constant $f$, which typically is a 
factor of $4\pi$ smaller than the scale of the new strong interactions.
An estimate for the corrections to the $hf\bar f$ couplings is~\cite{compositecoupling}
\beq
    g(f)/SM =  1 + (3-9)\% \cdot  \left( {\mbox{1\ TeV}\over f}\right)^2 \ .
\eeq{Compone}

In Little Higgs models, the Higgs boson couplings to $\gamma\gamma$ and $gg$
are modified by new contributions to the loop diagrams from the partners of
the top quark and the $W$ and $Z$ bosons.  These particles have masses in
the few-TeV range.  An estimate of the corrections is~\cite{LHcoupling}
\beqa
     g(g)/SM &=& 1+ (5-9)\% \CR
     g(\gamma)/SM &=& 1+ (5-6)\% 
\eeqa{LHmodels}

These results also illustrate the point made already in the previous section
that new physics corrections to the Higgs couplings can tweak any 
individual coupling independently of the others, so that a general, 
model-independent analysis of the couplings is needed.

After July 4, the issue of the precise values of the Higgs couplings 
has vaulted to the top of the list of the most important problems in 
high energy physics.   I have just argued that the level of precision
needed to address this problem is very high.  Can we get there?

\section{Poised}

During all of those years of waiting and hoping for the 
discovery of the Higgs
boson, many theorists and experimenters studied the prospects for 
new facilities that would dramatically improve our understanding of 
this particle.  We are poised to build them now.

\subsection{For the High Luminosity LHC}

Beyond the LHC at 14~TeV and $10^{34}$ luminosity, there is the 
High Luminosity LHC.   This planned upgrade of the LHC would begin 
its experimental data-taking in 2022.  The design gives a luminosity
greater than $10^{35}$/cm$^2$/sec, but also very challenging experimental
conditions with 200 pileup events per bunch crossing. This upgrade will
enable additional new particle searches, pushing the reach of the LHC
for gluinos and other strongly coupled new particles above 4~TeV~\cite{Fabiola}.
 
The HL-LHC initiative 
will produce huge statistics for Higgs analyses, a billion Higgs bosons
over 5 years.  But it will be very difficult to interpret these Higgs
events, or even to trigger on them.   Many important channels of
Higgs decay, especially the decays to $WW^*$ and $ZZ^*$, contain soft 
leptons.  For these channels, it is already a challenge to maintain
the trigger thresholds low enough to capture the events.  The study of 
Higgs decay to $b\bar b$ relies on excellent 2-jet mass resolution,
which will  be compromised by pileup.  The study of Higgs decay to
$\tau^+\tau^-$ and invisible modes relies on selection of
Vector Boson Fusion event using forward tagging jets.  The efficiency
for this selection will be compromised by large activity in the 
forward region $\eta > 2.5$. 

Thus, it is not obvious that there is any advantage for the study of 
Higgs couplings in increasing the ATLAS and CMS data samples from
300 fb$^{-1}$ to 3000 fb$^{-1}$.

However, there is a tremendous opportunity to be seized here.  The 
ATLAS and CMS collaborations are now studying ambitious detector 
modifications for the high luminosity era.  These include possible
pre-triggering or track-based triggering to improve the intelligence
of Level 1 event selection and new strategies to maintain performance
in the presence of pileup.   It must still be demonstrated that ATLAS
and CMS have useful capability for Higgs studies in the
high-luminosity era.  But I hope that the  members of the
collaborations will consider this a challenge that can be met.

\subsection{For an $\ee$ Higgs factory}

For the $Z$ and $W$ bosons, the discovery at hadron colliders was
followed by detailed precision study at the $\ee$ colliders SLC and
LEP.  The study of these particles in $\ee$ annihilation led to many
incisive experimental probes of the weak interactions, including
precision mass measurements and measurements of branching ratios and 
polarization asymmetries.
These experiments provide the foundation that we have today for our
understanding of the weak interactions.

There are equally good reasons to construct an $\ee$ collider to study
the Higgs boson.  In $\ee$ annihilation, Higgs boson production is
1\% of the total rate, rather than $10^{-10}$ as it is at hadron
colliders. This means that the various Higgs decay modes can all be
studied with minimum prejudice.  Decays of Higgs bosons to quarks, and
Higgs decays with 
hadronic decays of $W$ and $Z$, can be identified and used in
analyses.  This permits complete, unambiguous spin analysis of the Higgs 
boson couplings.  Decays of the Higgs to $c\bar c$ and $gg$ can be
identified
and distinguished from one another.  

 Most importantly, the process 
$\ee\to Zh$ allows the Higgs to be tagged by the presence 
of a recoiling $Z$
boson
at the correct energy in the center of mass system.  Using this
technique, it is possible to measure the absolute branching fractions
of Higgs decays.  Tagging of the Higgs also makes it possible to 
identify invisible Higgs decays, and also other possible exotic decay 
channels of the Higgs such as decay to long-lived particles.

Several technological solutions are now being proposed for the design
of lepton-collider Higgs factories, including synchrotrons~\cite{LEPthree},
linear colliders  with design energies up to
3~TeV~\cite{CLICreport},
and muon colliders with similar energy reach~\cite{MuonColl}.
However, among these solutions, the most compelling is the 
International Linear Collider.    The ILC has been extensively 
engineered over the past decade. 
 The ILC Technical Design Report is
now in preparation and should be completed before the end of the
year.  This is the one Higgs factory that can be proposed on the
correct time scale---immediately~\cite{Foster}.

\begin{figure}
\begin{center}
\includegraphics[width=6.0in]{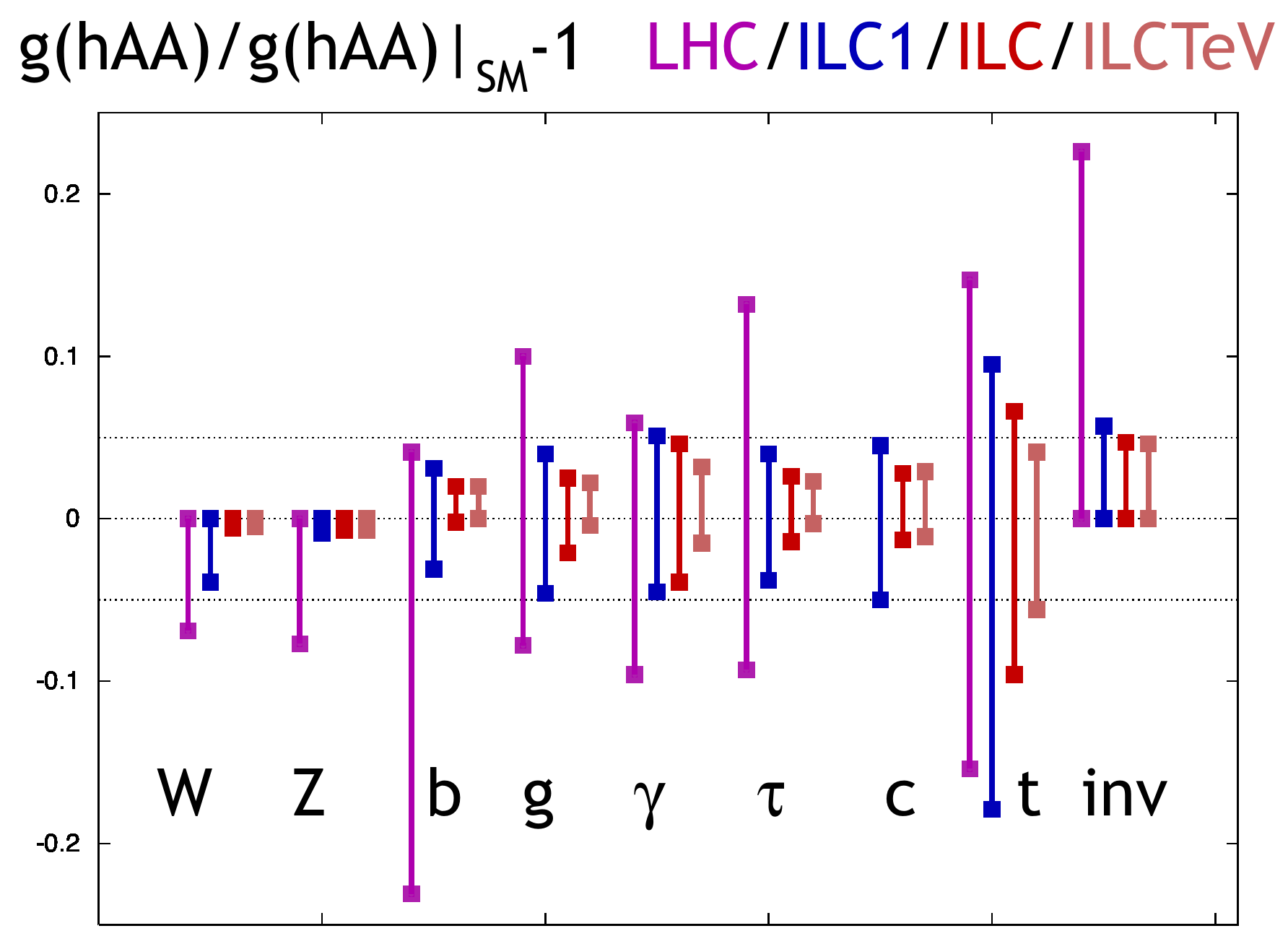}
\caption{Estimates of the accuracy that can be achieved in Higgs
  coupling measurements using a model-independent fit to LHC and ILC
  measurements, from \cite{MyHiggs}.
  The estimates are shown as a fraction of the predicted Standard
  Model value for the Higgs coupling constants.  The indicated
  horizontal lines
  represent 5\% deviations.  For the invisible
  Higgs decay,
 the quantity plotted is the square root of the branching fraction.
 The programs shown include (left to right for each entry)  LHC at
 14~TeV and 300~fb$^{-1}$,  ILC at 250~GeV and 250~fb${-1}$,
 ILC at 500~GeV and 500~fb${-1}$,  ILC at 1000~GeV and 1000~fb${-1}$.}
\label{fig:HiggsILC}
\end{center}
\end{figure}

 The capabilities of the ILC for
precision Higgs boson studies are very impressive.  The current
estimates are supported by full simulation detector studies with
realistic inclusion of the machine
environment~\cite{SIDLOI,ILDLOI,DBD}.
The improvements anticipated for the ILC over the estimates given
earlier for the 
LHC are shown in Fig.~\ref{fig:HiggsILC}~\cite{MyHiggs}.  These estimates correspond
to a nominal ILC program of 250 fb$^{-1}$ at 250 GeV, followed by
500~fb$^{-1}$ at 500 GeV, followed by 1000~fb$^{-1}$ at 1 TeV.   The
errors on statistics-limited modes such as $\tau^+\tau^-$ and
$\gamma\gamma$ would improve with longer running.   These estimates
imply that the ILC can meet the goals of a precision Higgs program, with errors
on individual couplings at the few-percent level in a
model-independent analysis.

\begin{figure}
\begin{center}
\includegraphics[width=6.0in]{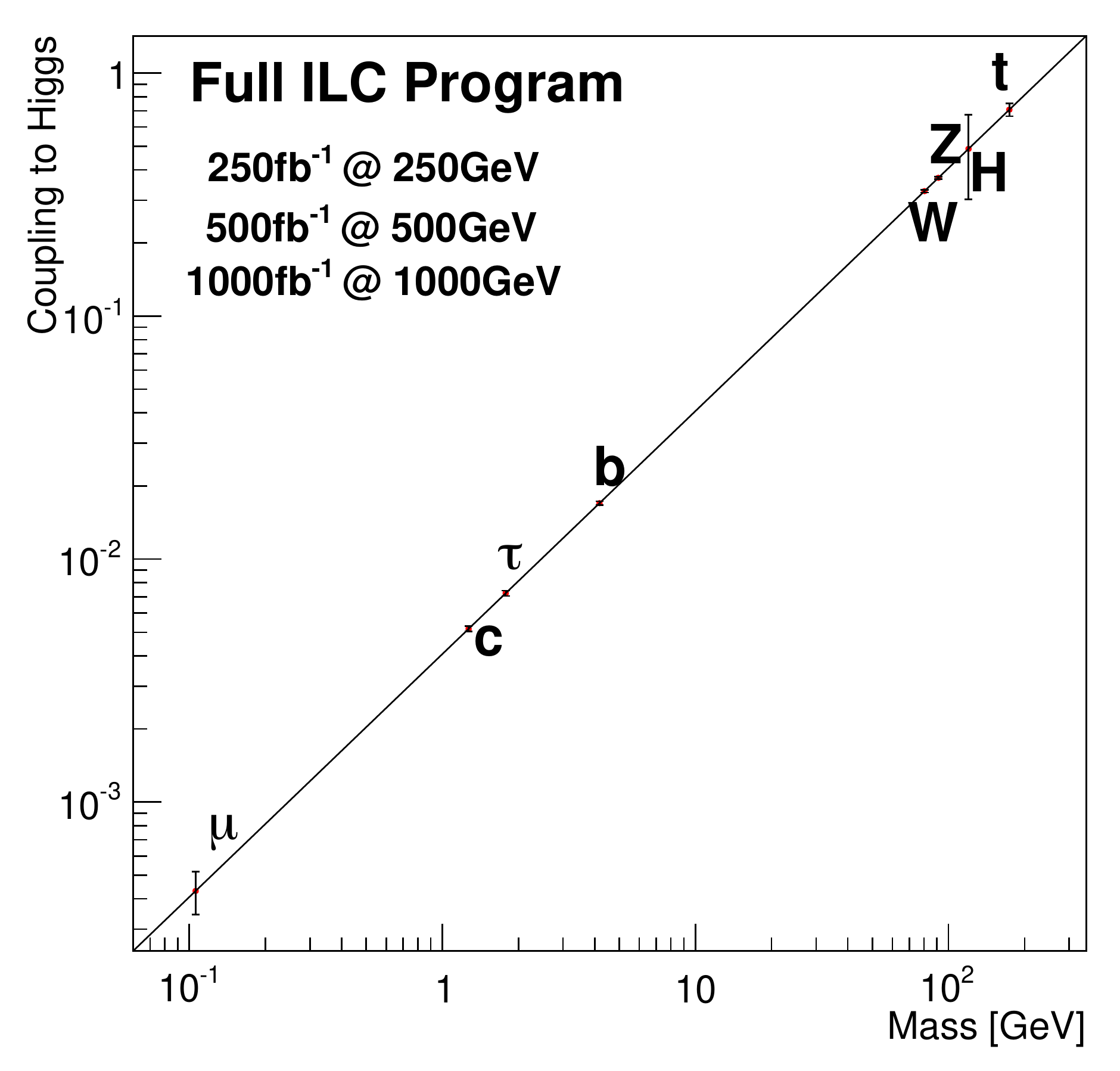}
\caption{Expected precision from the full ILC program of tests of the Standard Model relation
  that the Higgs couplings to each particle are proportional to 
its mass, from \cite{DBD}.   The measurements of the Higgs couplings
to $\mu$ and $t$ and the Higgs self-coupling require high energies;
the other measurements depend mainly on total integrated luminosity.}
\label{fig:linear}
\end{center}
\end{figure}

The expected precision of the test at the ILC of the proportionality of Higgs
couplings and mass is shown in Fig.~\ref{fig:linear}~\cite{DBD}.

 Over the years, much scorn has been poured on the ILC because its
design energy is ``only'' 500~GeV, extendable in a later stage to
1~TeV.  In the new era, though, those arguments have turned around
completely.    The first phase of LHC running has led to a discovery---the
Higgs boson.   The precision study of this particle could well be our 
only route to uncover new physics beyond the Standard Model. 

The ATLAS and CMS experiments have discovered no other new
particles.  At the moment, there is no case for new particles at
masses up to 1.5 TeV, calling for lepton collider  experiments at
3~TeV.  The LHC has eliminated many scenarios for physics beyond the
Standard Model that  seemed promising a few years ago. 
Remarkably, many of the scenarios for new physics that survive the
current LHC exclusions imply important experiments to be done in $\ee$ at 
500~GeV.  These include ``Natural Supersymmetry'' models in which the
lightest superparticles are Higgsinos with masses near 200~GeV~\cite{Baer} and composite Higgs 
models that call for a program of  precision
measurements on the top quark~\cite{LittleHiggs,RSHiggs}.    This
confirms the message that the new knowledge we have gained from the
LHC points to the importance of the ILC.

Whatever might be added from LHC discoveries later in this decade, 
the Higgs is there.
The ILC capabilities are perfectly matched to the needs of an
experimental program of precision measurements on the 
125~GeV Higgs
boson.  It is the right time, in direct response to the discovery, 
to call for  the construction of this machine.

\section{Conclusions}

The discovery of the Higgs boson implies an exciting and
program of beautiful observations to uncover the secrets that this
particle holds.   This program  will be
a major theme of high energy physics experimentation over the next ten
years.   It is likely to be full of mystery and surprises.

A new era of high energy physics, the Higgs era, has begun.  I am
awestruck at what has been accomplished in the first chapter of the
Higgs story, and I am impatient to see what the Higgs future may
bring. We are ready to move forward to make these discoveries.

\Acknowledgements

I thank Louis Fayard for the invitation to lecture at Higgs Hunting 
2012.   I am grateful also to Fayard,  Zoltan Kunszt,
Joe Incandela, Marcello Manelli, Matt Strassler, and many other participants
in the workshop for sharing their insights.  I thank Tim Barklow, Kyle Cranmer,
Keisuke Fujii, Howard Haber, Heather Logan, Francois Richard, Tim Tait,
and many other colleagues 
with whom I have discussed precision Higgs physics in preparation for this 
report.   This work was supported by the US Department of
Energy under contract  DE--AC02--76SF00515.

\end{document}